# On the Philosophical Foundations of Physics: An Experimentalist's Perspective

Stephen Boughn[†]
Department of Physics, Princeton University and
Departments of Astronomy and Physics, Haverford College

**Contents**

**Preface**

Since I first became enthralled with physics as a teenager, I've been intrigued by the philosophical aspects of the discipline. I'm certainly not alone in being fascinated with such heady notions as the absence of absolute simultaneity, the twin paradox, curved space-time, wave particle duality, the collapse of the wave function, Schrödinger's cat, and more recently dark matter and dark energy. I was chagrined when I read Richard Feynman's disparaging words about the relation of philosophy and physics[1]. On the other hand, the more I read Feynman, especially his popular articles, the more I realized that he was, indeed, a very philosophical physicist. Perhaps the footnote on the previous page indicates that he had more of a problem with (some) philosophers than with the philosophical conundrums that have so beguiled me.

---

[†] sboughn@haverford.edu
[1] Near the end of the 7th Messenger Lecture at Cornell in 1964, Feynman opined that fundamental physics will inevitably come to an end "because either all is known, or it gets very dull" (i.e., it becomes too difficult to make significant progress). Then "The philosophers who are always on the outside making stupid remarks will be able to close in, because we cannot push them away..."

As I approached the end of my career as an experimental physicist and observational astronomer (I'm now retired), I decided to return to these philosophical matters, some of which were still perturbing me, to see if I could finally make enough sense of them to quiet my discomfort. I've more or less succeeded in this quest in large part, I believe, because of my experimentalist background and a concomitant proclivity for pragmatic explanation. I've written several papers that include many of the ruminations in the present paper and these are listed below in the References.[2] I suspect that many readers of this paper will consider my practical approach to be rather pedestrian[3] but I am, after all, a pragmatic experimentalist. In any case, I here offer my thoughts *on the philosophical foundations of physics*.

In preparation, I've spent many (mostly) enjoyable hours reading, and in some cases re-reading, papers and books relevant to the philosophical foundations of physics. I was somewhat taken aback by the fact that several different aspects of my "original" thoughts on the philosophy of physics indeed appear scattered throughout many of these works. In addition, I suspect that the many of the remainder of my thoughts can be found in papers, which have likewise been influenced by previous works. I know this from my recent readings of books by Mach (1893), Carnap (1966), Jammer (1974), and even William James (1908). At first I thought how clever I was to arrive at some of the same conclusions as these notable philosophers. But, alas, it seems far more likely that I had simply stumbled upon some of their thoughts that were imbedded in other works. From the title of this paper, it is clear that I supposed my experimentalist's view of the philosophy of physics might provide a refreshingly different perspective. Considering that Mach, who was arguably the first modern philosopher of physics, was an experimentalist, even this notion is suspect. One paper that has been foremost in my mind ever since I first read it in graduate school is "The Copenhagen Interpretation" by the theorist Henry Stapp (1972). This paper, to which I will liberally refer, certainly has had a profound effect on my view of quantum mechanics in particular and of physics in general.

---

[2] I am not in the habit of citing so many of my own papers; however, I do so here because they are sources of a great deal of the present paper.

[3] Indeed, the title of one of the papers listed in the References is "A Pedestrian Approach to the Measurement Problem in Quantum Mechanics".

While I will make reference to the abovementioned works as well as to others, I make no attempt to include a comprehensive history of the philosophy of physics or even to carefully assess the works of others in relation to the views presented here. I'm pleased that such an approach was given tacit approval by James (1908) who proclaimed

> The abuse of technicality is seen in the infrequency with which, in philosophical literature, metaphysical questions are discussed directly and on their own merits. Almost always they are handled as if through a heavy woolen curtain, the veil of previous philosophers' opinions. Alternatives are wrapped in proper names, as if it were indecent for a truth to go naked….You must tie your opinions to Aristotle's or Spinoza's; you must define it by its distance from Kant's; you must refute your rival's view by identifying it with Protagoras's. Thus does all spontaneity of thought, all freshness of conception, get destroyed.

It may seem absurd that a relatively short paper could possibly present a meaningful account of the philosophical foundations of physics. Carnap's 300 page book with a similar title treated only a restricted number of fundamental problems. Jammer's text, *The philosophy of quantum mechanics*, is 500 pages as is Mach's *The science of mechanics: a critical and historical expositions of its principles*. How can I possibly expect to say anything meaningful in the less than 50 pages of the present essay? I can hear my friend and graduate school classmate, physicist and philosopher Peter Pesic, whispering over my shoulder, "This is an interesting piece. Why don't you delve more deeply into the matter and then write a book about it."[4] Well, I neither want to write a book nor do I want to delve more deeply into the subject. My purpose here is to sketch a worldview with which one might be able to approach fundamental philosophical and interpretational problems. In fact, it might be an advantage to avoid the depth and precision that would limit flexibility in dealing with the philosophical conundrums I seek to resolve. In his correspondence with Stapp (1972), Heisenberg pointed to an analogous vagueness in the Copenhagen interpretation: "Besides that it may be a point in the Copenhagen interpretation that its language has a certain degree of vagueness, and I doubt whether it can become clearer by trying to avoid this vagueness." From what follows you'll see why I view this attitude as a strength and not as a weakness.

---

[4] He actually did make such a suggestion regarding a previous paper ("A Quantum Story"), one of those listed in the References.

In the next section, I'll begin with my resolution to one of the conundrums of the measurement problem in quantum mechanics, the quantum/classical divide. I say "my" resolution but in large part it is informed by Stapp's paper as well as by the writings of Niels Bohr, Eugene Wigner, Wendell Furry, Freeman Dyson (2015), and I'm sure of many others. The pragmatic resolution of this problem will provide a useful example that will help situate my experimentalist's approach to the philosophical foundation of physics, on which I'll elaborate in the following section (Section 2). Section 3 will address several other well known philosophical puzzles in physics in the context of my pragmatic experimentalist's approach.

1. THE MEASURMENT PROBLEM AND THE QUANTUM-CLASSICAL DIVIDE

One of the most important aspects of the measurement problem in quantum mechanics has to do with the description of the measuring apparatus. It is generally accepted that the Schrödinger wave function describes the dynamic evolution of a quantum system. It is to be interpreted statistically, that is, the wave function indicates the probability of the occurrence of the result of a specific measurement. On the other hand, one expects quantum mechanics to be a universal theory so that the experimental apparatus should also be subject to quantum mechanical laws. If so, then one is left with a probabilistic description of the measuring apparatus. This presents a problem. The interpretation of the original wave function requires that a specific outcome eventually be achieved but if the experimental apparatus is described by another probabilistic wave function, no such outcome can be identified. So it seems we need a $2^{nd}$ measuring apparatus to exam the original apparatus but the laws of quantum mechanics also govern this second apparatus so it too must be described by a probability function. Then a $3^{rd}$ apparatus is required and so on *ad infinitum*. Heisenberg expressed this in the extreme, "One may treat the whole world as one [quantum] mechanical system, but then only a mathematical problem remains while access to observation is closed off." (Quoted in Schlosshauer and Camilleri 2011) One of the early resolutions to this dilemma, and one that is still entertained today, is the postulate of *state reduction* or *collapse of the wave function*. It's generally accepted that this notion requires new physics that has yet to be discovered. In Section 3.3, I'll discuss why wave function collapse is neither required nor particularly well motivated.

The standard textbook resolution of the measurement problem is that offered by Niels Bohr and Werner Heisenberg and is generally referred to as the Copenhagen interpretation. According to the Copenhagen interpretation, or so the standard argument goes, the act of measurement or observation must be described in terms of classical physics. Because classical physics is not (generally) couched in probabilistic terms, the classical measurement of a quantum system yields the specific result required by the probabilistic interpretation of the wave function. While possibly resolving the predicament, this explanation immediately raises three related questions: 1) Why is it necessary to revert to classical physics in order to understand quantum mechanics? 2) When and where does the classical measurement occur, i.e., where is the quantum-classical divide? and 3) How does one describe the physical interaction across this divide? These are questions that have vexed physicists and philosophers since the beginning of quantum mechanics.

Hugh Everett, the originator of the *many worlds interpretation* of quantum mechanics, expressed a strong objection to the notion of including classical physics in the interpretation of quantum mechanics. In his 1956 PhD thesis (DeWitt & Graham 1973), he argued:

> Another objectionable feature of this position [Bohr's Copenhagen interpretation] is its strong reliance upon the classical level from the outset, which precludes any possibility of explaining this level on the basis of an underlying quantum theory. (The deduction of classical phenomena from quantum theory is impossible simply because no meaningful statements can be made without pre-existing classical apparatus to serve as a reference frame.) This interpretation suffers from the dualism of adhering to a "reality" concept (i.e., the possibility of objective description) on the classical level but renouncing the same in the quantum domain.

So why didn't Bohr seem to worry about the unseemly merger of classical and quantum formalisms? I suspect the answer is because he didn't actually consider that any such merger is required. While Bohr endeavored to be extremely careful in expressing his ideas, his prose is often obscure. However, consider his following brief description of a measurement (Bohr 1963, p. 3):

> The decisive point is to recognize that the description of the experimental arrangement and the recordings of observations must be given in plain language, suitably refined by the usual terminology. This is a simple logical demand, since by the word 'experiment' we can only mean a procedure regarding which we are able to communicate to others what we have done and what we have learnt.

Nowhere in this description does he refer to classical physics. Stapp (1972) chose to emphasize this pragmatic view by using the word "specifications":

> Specifications are what architects and builders, and mechanics and machinists, use to communicate to one another conditions on the concrete social realities or actualities that bind their lives together. It is hard to think of a theoretical concept that could have a more objective meaning. Specifications are described in technical jargon that is an extension of everyday language. This language may incorporate concepts from classical physics. But this fact in no way implies that these concepts are valid beyond the realm in which they are used by technicians.

The claim is that descriptions of experiments are invariably given in terms of operational prescriptions or specifications that can be communicated to the technicians, engineers, and the physics community at large. Such operational prescriptions are not part and parcel of classical mechanics. They are not couched in terms of point particles, rigid solid bodies, Newton's laws or Hamilton-Jacobi theory but rather are part of Bohr's "procedure regarding which we are able to communicate to others what we have done and what we have learnt." Camilleri and Schlosshauer (2015) point out, "Bohr's doctrine of classical concepts is not primarily an interpretation of quantum mechanics (although it certainly bears on it), but rather is an attempt by Bohr to elaborate an epistemology of experiment." So it seems to me that Everett's (and others') assertion that Bohr insisted on merging a quantum system with a classical measuring apparatus is a "straw man" that has little to do with Bohr's interpretation of quantum mechanics. In fact, one can find several passages in Bohr's writings where he claims that the physics of experimental apparatus is certainly describable by the formalism of quantum mechanics (Bohr 2010, Camilleri & Schlosshauer 2015). But that description has little to do with the "epistemology of experiment".

In some sense, classical physics has precisely the same problem in that experiments are not described by the formalism of the theory but by the same operational specifications to which Stapp refers (Boughn and Reginatto 2019). This point has certainly not gone unnoticed by those trying to come to grips with the quantum/classical divide. At a 1962 conference on the foundations of quantum mechanics, Wendell Furry explained (Furry 1962)

> So that in quantum theory we have something not really worse than we had in classical theory. In both theories you don't say what you do when you make a measurement, what the process is. But in quantum theory we have our attention focused on this situation. And we do become uncomfortable about it, because we have to talk about the effects of the measurement on the systems....I am asking for something that the formalism doesn't contain, finally when you describe a measurement. Now, classical theory doesn't contain any description of measurement. It doesn't contain anywhere near as much theory of measurement as we have here [in quantum mechanics]. There is a gap in the quantum mechanical theory of measurement. In classical theory there is practically no theory of measurement at all, as far as I know.

At that same conference Eugene Wigner put it like this. (Wigner 1962)

> Now, how does the experimentalist know that this apparatus will measure for him the position? "Oh", you say, "he observed that apparatus. He looked at it." Well that means that he carried out a measurement on it. How did he know that the apparatus with which he carried out that measurement will tell him the properties of the apparatus? Fundamentally, this is again a chain which has no beginning [end]. And at the end we have to say, "We learned that as children how to judge what is around us." And there is no way to do this scientifically. The fact that in quantum mechanics we try to analyze the measurement process only brought this home to us that much sharply.

Physicists have long since become comfortable with the relation between theory and measurement in classical physics, so the quantum case should not be viewed as particularly vexing.

So it seems that experiments and observations are essential in giving meaning to both classical and quantum mechanics. As Heisenberg pointed out, without observations the theoretical formalism provides us with no more than mathematical problems that are disconnected with the outside world. Quantum and classical formalisms provide no instructions on how a measurement is to be performed. Physical theories are compact mathematical models together with a set of rules for interpreting the formalism in terms of observations and experiments but are silent on how to perform the latter. On the other hand, experimental physics is an immense archive of descriptions of instruments and procedures partly based on classical and quantum theory, to be sure, but also on previous experiments, phenomenology, conventional wisdom, past experience, physical intuition, and with a large emphasis on the importance of calibration, which obviates the need for the complete understanding of an experimental apparatus. In addition, the entries in these tomes have to be translated into Stapp's specifications so that experimental physicists, engineers, and technicians can create the apparatus necessary to make and then interpret a measurement. These specifications are given in Bohr's "plain language", which is not the formal language of classical mechanics. The relations and theoretical constructs of the formalism of classical mechanics are not part of the language used to communicate the specifications needed by engineers and technicians to build and carry out experiments.

It seems likely that the measurement problem in quantum mechanics has been exacerbated by the lack of experimental expertise possessed by many of those scholars who hold forth on the dilemma of quantum measurement and this situation has contributed to the intractability of the problem. This is not to say that the fuzzy connection between theory and experiment has not been grasped by many physicists. In his paper, "Physics and Reality" Einstein (1936) mused,

> The connection of the elementary concepts of every day thinking with complexes of sense experiences can only be comprehended intuitively and it is unadaptable to scientifically logical fixation. The totality of these connections, - none of which is expressible in notional terms, - is the only thing which differentiates the great building which is science from a logical but empty scheme of concepts. By means of these connections, the purely notional theorems of science become statements about complexes of sense experiences... Physics constitutes a logical system of thought...The justification (truth content) of the system rests in the proof of usefulness of the resulting theorems on the basis of sense experiences[5], where the relations of the latter to the former can only be comprehended intuitively.

It should be obvious why the resolution of the measurement problem offered by the Copenhagen interpretation appeals to my experimentalist's pragmatic view of the world. Even though it was conjured by two theoreticians, Bohr and Heisenberg, it's also obvious why theoretical physicists might be less sanguine about it. In any case, I hope that the above treatment will serve to set the stage for my experimentalist's view of the philosophical foundations of physics, to which I now turn.

---

[5] There is a certain irony in this statement. It seems to be entirely consistent with a pragmatic philosophy and by extension with Bohr's view as we shall see in the next section.

2.  PRAGMATISM AND THE PHILSOPHICAL FOUNDATIONS OF PHYSICS

While I found most of my readings in the philosophical foundations of physics to be interesting, very few resonated with me.  Most were quite detailed and theoretical in a way that didn't capture my view of the totality of physics, theory AND experiment.  For example von Neumann's treatment of a quantum mechanical experiment, although still considered relevant today, is nearly unrecognizable to me as the description of an experiment.  In his 1932 book, *Mathematische Grundlagen der Quantenmechanik* (von Neumann 1955), von Neumann provides a quantum mechanical treatment of an experimental apparatus by ascribing to it a wave function $\xi(g_n)$ characterized by a pointer value $g_n$ that is associated with the eigenvalue $A_n$ of some quantum operator **A** . The wave function of the combined system (quantum system plus experimental apparatus) is given by $\Psi = \sum_n \xi(g_n)\varphi_n$ where $\varphi_n$ are the eigenfunctions of **A** associated with the quantum system to be measured. It is assumed that the pointer value can be accurately determined from an examination of the apparatus by any observer (although, such an examination presumably takes place outside the confines of the quantum mechanics).  While one might argue that the identification of the wave function $\xi(g_n)$ of an experimental apparatus is possible in principle, the practicality of such an identification is clearly absurd.[6]  In addition, the notion that $\xi(g_n)$ could possibly capture all the nuances of the description of an experiment, including Stapp's operational specifications, is ludicrous.

I think my pragmatist's view of the foundations of physics began to crystallize while pondering the redoubtable 1935 paper by Einstein, Podolsky, and Rosen (EPR)[7].  I first read the EPR paper when I was in graduate school and was mystified. I had assumed the paper would present an argument concerning the self-consistency of quantum mechanics or lack there of.  Instead, it seemed to be less about physics and more about the philosophical issue of what constitutes reality. The conclusion of EPR was quite simply that quantum mechanics "does not provide a complete [true] description of the physical reality". It was only recently that I began to worry about, not the veracity of EPR's claim, but about the very meaning of "a complete and true description of physical reality". Just what do these words mean? I think I know what one means by "a complete and true description of a house". It means that one is provided with a complete set of photographs, architectural drawings, list of construction materials, etc. However, these items and the person reviewing them are certainly external to what we mean by "the house". On the other hand, I'm not so sure I know what one means by a complete and true description of the physical reality. In this case, both the description and the beings who generate and review the description are surely part of physical reality. If so, a complete description must include the description itself and the people generating and reviewing it. I have no idea how this might be accomplished. It would seem to require that the recipient of such "a complete description of the physical reality" be an entity external to physical reality, which would place the whole discussion in the realm of religion or mythology and well outside that of physics.[8] In a recent essay (Boughn 2019a) I observed that whenever the concept of *reality* is raised in the context of interpretations of quantum mechanics it is a warning to me that I'm about to be led down a rabbit hole and, therefore, suggested that in physics discourse the term "reality" should be avoided at all costs. Even so, any conversation about the philosophical foundations of physics must eventually deal explicitly with the notions of *reality* and *truth*.

---

[6] For example, assigning a wave function to the 7000 tonne ATLAS detector at the Large Hadron Collider ( http://img804.imageshack.us/img804/8259/sl2407020161.jpg ), with which the Higgs boson was recently discovered, is inconceivable to me.

[7] The EPR paradox, Bell's theorem, and spooky action at a distance are the subject of Section 3.5.

      During the last millennium, philosophers have discussed the *real world* in an extraordinary number of ways: logical positivism, logical empiricism, radical empiricism, pragmatism, pluralism, realism, rationalism, materialism, monism, idealism, dualism, pantheism, just to name a few. I sometimes think that such labels have been created and applied to the work of individual philosophers in an effort to pigeonhole them for the purpose of criticizing their thoughts.[9] I've often found that I agree with some of the ideas presented in the context of many of these philosophical theories while objecting to other ideas in these same theories. Luckily, I'm following Williams James's advice (given in the preface) and, therefore, will resist categorizing my philosophy in this way. Nor will I attempt to place it in

---

[8] I appreciate the irony that my objection to EPR's conclusion is analogous to those who object to the Copenhagen interpretation's contention that the experimental apparatus and observer cannot be included in the formal quantum mechanical description of the complete system.

[9] I suppose that such labels also find use in generating exams for university students. I'm also dubious about this practice.

the context of the theories of others.

So how do I characterize the *real world?* I take it as a matter of faith that the real world exists. Why? It's quite simple. Every morning when I wake up, I experience the real world. I'll rarely use the term "reality" in the remainder of this paper and when I do it will be in the same context; reality is what I experience. The question then becomes, how do people make sense of these experiences. The answer is again quite simple. They make up stories about their experiences. The purposes of these stories are several: to facilitate communicating experiences to others; to draw comparisons to other experiences; to predict future experiences (e.g., sunrises and sunsets); to motivate actions that (presumably) bring about changes to the "real world", changes that lead to predictable future experiences; in short to make sense of experiences and, by extension, to make sense of the real world. Do these stories provide "true" descriptions of the real world? Of course not; they are stories, human inventions created to help us deal with our experiences (see the Einstein quote in Section 1). On the other hand, our stories certainly do provide insight into the natural world and thereby help us to understand it. But in the end, the only direct connection our stories have is to our experiences. The relation of such stories to physics is simple. Our formal, and usually mathematical, physical theories are, in the end, just stories that we use to make sense of our experiences in the ways just described.

Of course, some (many) of these stories will be simply wrong. What we value are "true" stories. So what makes a story or a physical theory *true*. The sense I get from the EPR paper is that a true theory must provide a complete description of reality, which they characterize as having an accurate one-to-one correspondence between individual theoretical states and all aspects of reality. This probably goes beyond what most physicists mean by truth. Many of us would probably be content with a provisional truth characterized by the confirmation of predictions of the theory; although, this may not go far enough. I suspect that the realists among us might hold onto the Einsteinian ideal that a true statement is one that accurately describes some aspect of the real world, whether or not that truth is ultimately demonstrable. As you might expect, since I don't subscribe to a reality that can ever be accurately described, I find this notion of truth to be equally vacuous.

At this point, I unashamedly fall back on James's pragmatic definition of truth as the identification of the degree of *utility* of the theory. A theory or story is true to the extent that it is useful in making sense of our experiences.  In the absence of direct knowledge of the physical world, this is the best we can do.  In the case of physics, the experimental or observational confirmation of the predictions of a theory is, perhaps, the most common expression of the usefulness of a theory and therefore of the truth of a theory.  However, the degree to which a physical theory suggests connections to other theories or to extensions of the current theory is also a measure of the usefulness (truth) of the theory.  Certainly the ease with which predictions can be generated from a theory and, I suppose, even the elegance of a theory can be considered measures of its utility.

The previous two paragraphs constitute a bare-bones account of my pragmatic view of the philosophical foundations of physics.  "Bare-bones" indeed!  One might rightfully complain that I've given so little detail as to render my worldview of little use to either philosophy or physics.  Even my allusion to James's account of *truth* is abbreviated in the extreme.  He felt it necessary to write many dozens of pages on the topic.  To be sure, James delves deeply into the subject, which by necessity involves such considerations as experiences, feelings, consciousness, cognition, cognitive relations, senses, knowing, mental content, common percepts, ideas, and of course reality, among others.  I am certainly not criticizing his analysis but would point out that his efforts had much more to do with psychology and neuroscience than with physics.  I'm sure there are many reasons why one might pursue such an elaborate model, for example, a desire for a complete description in the sense as expressed in the EPR paper, an attempt to include a direct path to the resolution of any philosophical problem that might be raised, or even an effort to differentiate one's philosophy from all others.  Whatever the reason, such detailed worldviews often leave little room for my perspective of "the fuzzy connection between theory and experiment" nor even of Einstein's contention that "complexes of sense experiences can only be comprehended intuitively".  As I wrote in the preface, I feel that "it might be an advantage to avoid the depth and precision that would limit flexibility in dealing with philosophical conundrums."  This should be already apparent in the resolution of the quantum/classical divide that was presented in Section 1 and I hope that the cases dealt with in the following section will provide further examples of the utility of my bare-bones pragmatic worldview.  On the other hand, the vagueness of this philosophy leaves open the possibility of entertaining novel ways of treating specific physical and philosophical problems.

There are two corollaries to my pragmatic philosophy of physics. Many years ago, I came upon the Einstein quote, "It is the theory which decides what we can observe." As you can imagine, this declaration was an irritation to my experimentalist temperament. John Bell repeated Einstein's thought in a 1990 paper entitled, "Against Measurement".[10] As far as I can tell, Einstein's utterance came during a 1926 conversation with Werner Heisenberg. It was Heisenberg's first meeting with Einstein and he relates the details of it in *Physics and Beyond* (Heisenberg 1971). Einstein was expressing his disagreement with Heisenberg's quantum mechanics and its refusal to even consider the orbits of electrons within an atom. In his defense, Heisenberg made the comparison to Einstein's rejection of the concept of absolute time and simultaneity in special relativity.[11] Their extended conversation somewhat soothed my irritation. Given Einstein's argument in the subsequent EPR paper, I will take the liberty to re-express his sentiment. According to Einstein, every observation must be describable by an unambiguous statement in the physical theory. Okay, but let me turn this statement around with the claim that *every meaningful statement in a physical theory must be directly (or perhaps indirectly) related to some observation*. This is the *first corollary* to my pragmatic philosophy and I will use it to judge whether or not statements in the context of a physical theory are meaningful. The very first example in the next section deals with the notion of simultaneity in special relativity.[12]

The second, related corollary has to do with the status of theoretical constructs, such as the Schrödinger wave function. There is a tendency to elevate the constructs of highly successful theories to the status of "real" entities and subsequently to posit questions about them as if they resided in the natural world. Of course, the structure of a theory often provides insight into how to extend that theory or even on how to replace it with its successor. But that's a far cry from endowing a theoretical construct with reality in the sense that Einstein apparently required of a quantum state (if quantum mechanics were to provide "a complete description of the physical reality"). In an effort to resist the tendency to attribute physical reality to theoretical constructs of successful models, perhaps we should apply to such constructs an aphorism analogous to that offered by Stapp for the description of a measurement (see Section 1), something like, *the constructs of a theoretical model may incorporate concepts of classical (and everyday) language; however, this fact in no way implies that these concepts are valid beyond the realm in which they are used by physicists*. In other words, theorists must declare the meaning of such constructs and not try to discover their meanings as if they were part of the natural world. This is the *second corollary* to my pragmatic philosophy.

---

[10] Bell's paper similarly irritated me.

[11] Einstein retort to Heisenberg : "Possibly I did use this kind of reasoning, but it is nonsense all the same." (Heisenberg 1971) When Philipp Frank made a similar plea, Einstein's response was "A good joke should not be repeated too often." (Isaacson 2007)

[12] Heisenberg also used this example in his 1927 paper on the indeterminacy relation (see Section 3.2).

The legitimate application of this *second corollary* is not always immediately apparent. For example, one might consider negative energy solutions as outside the realm of the intended use of the Dirac equation. Nevertheless, these solutions led to the 1931 prediction of positrons, which were detected two years later. I prefer to think that Dirac happened upon a model of the electron that was also valid for the anti-electron (positron) should it exist.[13] On the other hand, pragmatist that I am, this inherent fuzziness of my *second corollary* is not particularly bothersome to me.

Before leaving this section, I'll offer my views on what is probably the more conventional philosophy of physics that most practicing physicists have at least somewhere in the backs of their minds.[14] It seems to me that most physicists ascribe to the notion that there is a real physical world and that world is governed by a set of (presumably mathematical) physical laws. The job of physicists is then to *discover* these laws in the same sense that explorers discover new lands. This view is certainly consistent with that of Einstein's, whether or not one accepts that a single such law provides a complete description of the real world (a theory of everything). This view is to be contrasted with my pragmatic view that the job of physicists is to investigate phenomena, i.e., accumulate new experiences, and then create mathematical models (stories) that provide true (useful) descriptions of those experiences. The former view portrays the physicist as an ingenious investigator while the latter as a creative genius. As far as I'm concerned these two characterizations are equally praiseworthy. On the other hand, one might well argue that adopting either of these philosophies is of no practical consequence. While contemplating foundational questions might be compelling (to some), "doing physics" doesn't seem to require that such questions be addressed. This attitude can be summed up by David Mermin's assessment of the Copenhagen interpretation: "If I were forced to sum up in one sentence what the Copenhagen interpretation says to me, it would be `Shut up and calculate!'" (Mermin 1989) I'm certainly sympathetic with this view; however, when it comes to dealing with some of the paradoxes and conundrums of physics, one's philosophical viewpoint is crucial as I hope the following accounts in Section 3 will reveal.

---

[13] The negative energy solutions of the Dirac equation don't violate my *first corollary* because Dirac was able to point to their observable consequences.

[14] I hasten to add that many physicists undoubtedly forego such metaphysical deliberations and are content to get on with the business of doing physics.

It is an interesting historical question as to why most scientists, physicists in particular, ascribe to the former "ingenious investigator" perspective rather than to that of the "creative genius". I'm certainly no historian of physics but let me just toss out a possibility that has intrigued me for the last 40 years.[15] The history of science in the Renaissance was marked by the rediscovery of ancient texts. Among these were the Hermetica, the writings of the mythical figure Hermes Trismegistus, who was supposedly a contemporary of Moses. "An account of how Hermes Trismegistus received the name "Thrice Great" is derived from the _Emerald Tablet of Hermes Trismegistus_,[16] wherein it is stated that he knew the three parts of the wisdom of the whole universe" (Hermeticism, Wikipedia). This picture of scholars mining ancient documents to find the wisdom therein strikes me as metaphor for the modern view of physics as a quest to find physical laws writ large in Nature. For Newton, it was perhaps more than a metaphor. He was convinced that Pythagoras knew the inverse square law of gravitation and that Moses had raised the possibility that some of the bodies God created were much larger than the earth and, perhaps, were habitable (Iliffe 2017). Much more recently, P. A. M. Dirac emphasized the importance of searching for fundamental physical truths in the very mathematics eventually used in formal models of these truths.[17] As he expressed it (Dirac 1963)

> It may well be that the next advance in physics will come along
> these lines: people first discovering the equations and then
> needing a few years of development in order to find the physical
> idea behind these equations. My own belief is that this is a more likely line of progress than trying to guess at physical pictures.

---

[15] I have a distinct memory of a 1976 dinner conversation with an old friend, Paul Boynton (University of Washington), who introduced me to this notion. At least this is they way I remember it.

[16] A translation of the _Emerald Tablet_ can be found in Isaac Newton's papers.

[17] Max Tegmark has expressed this in the extreme by proposing that the real world is precisely its mathematical structure (Tegmark 2008).

While Dirac's view may be a bit extreme, it strikes me that many (most?) theorists seem to proceed as if they are searching for the mathematical formalism that represents the laws of physics as if those mathematical laws have been preordained by Nature herself. This is certainly at odds with my pragmatist's view that physicists themselves create theories solely for the purpose of making sense of our experiences.

## 3. CONUNDRUMS IN THE PHILOSOPHICAL FOUNDATIONS OF PHYSICS

The following are examples[18] of how philosophic pragmatism (or at least my version of it) can be used to resolve well-known foundational dilemmas in physics. The first example, the lack of absolute simultaneity in special relativity, was resolved long ago but serves as a useful illustration of the pragmatic argument. Also, there continue to be misleading proclamations about special relativity, especially regarding the twin paradox.

### 3.1 SIMULTANEITY, TIME DILATION, AND THE TWIN PARADOX

One of the main lessons I learned from special relativity was to pay strict attention to the degree to which theoretical quantities can be measured, which clearly aligns with my experimentalist background. Einstein based his 1905 analysis on the empirical relation that the speed of light was a constant, independent of the motion of an inertial observer. He then rendered this ostensibly absurd notion to be self-consistent by prescribing *time* to be simply that quantity which is measured by an observer with an ideal clock and not some physical (or metaphysical) entity that flows through and determines the evolution of the universe. In short, Einstein's position was entirely consistent with the *1st corollary* offered above. He then proceeded, with the aid of a light clock, to demonstrate that the notion of the *simultaneity* of two events is not absolute but rather relative to particular observers and their clocks. This was the essence of Heisenberg's 1926 response to Einstein's complaint about quantum mechanics (see Section 2).

---

[18] The quantum/classical divide of Section 1 should be considered as another example of how one of the aspects of the quantum measurement problem can be successfully resolved in the context of a pragmatic philosophy.

Once one accepts the consequences of the relative nature of simultaneity, most of the other perplexing aspects of special relativity become transparent. The twin paradox is a prime example. Nevertheless, while this "paradox" was resolved long ago, I'm amazed at how many of my students (and occasionally colleagues) still seem to be mystified by it. One of the standard resolutions of the twin paradox is the realization that the experiences of the two twins are not the same; the one who has experienced acceleration ages less and the other twin who has not experienced acceleration has aged more. Unfortunately, this leads the novice student of relativity to conclude that the differential aging is a direct result of the acceleration of the younger twin and even more perplexing it has led some physicists to claim that the twin paradox can only be resolved within the general theory of relativity.[19] To help dispel these notions I conjured a twin paradox scenario wherein the two twins have identical experiences, acceleration and all, and yet one twin ends up younger than the other (Boughn 1989). Even so, many students of physics are not satisfied. Not content with the special relativistic prediction of the behavior of clocks, they want to know "why" the acceleration causes one of the twins to age less and "when" the extra aging of the stay-at-home twin occurs. These seem like meaningful questions; however, neither of them can be identified with an observation and, therefore, they are rendered meaningless by the *1st corollary* of my pragmatic philosophy.[20] My claim is that all of the apparent paradoxes of special relativity involve theoretical statements that have no connections to observations and, therefore, can be similarly dismissed. I leave it as an exercise for the reader to verify this claim.

Most of us consider special relativity to be a complete, self-consistent formalism that is capable of predicting the outcomes of all observations of electromagnetic phenomena that don't involve other physics such as the quantum properties of atoms. This might lead one to suspect that Einstein's ideal clocks are fictions whereas a pragmatic experimentalist must deal with real, physical clocks. This question must be addressed before one can fully analyze scenarios like the twin paradox. I have argued (Boughn 2013) that the effects of acceleration $a$ on real atomic clocks are of the order of $al/c^2$ where $l$ is some characteristic length for the 18clock and $c$ is the speed of light.[21] The characteristic length for an atomic clock is undoubtedly larger than the size of an atom but is surely no larger than the physical size of the clock apparatus, i.e., $l < \sim 1m$. For an acceleration equal to that of gravity on the surface of the earth and $l < 1m$, we have $al/c^2 < 10^{-16}$, comparable to the accuracy of the best atomic clocks. So for current time determination and reasonable accelerations, our best clocks are good approximations to Einstein's ideal clocks. Nevertheless, this exercise serves to emphasize the importance of experiment to the foundations of even special relativity.

---

[19] To the contrary, special relativity is quite capable of dealing with accelerated clocks and observers.

[20] To be sure, it is possible to specify measurements that seem to answer the question of when the extra aging occurs; however, such measurements are altogether arbitrary and certainly not unique.

3.2 HEISENBERG'S INDETERMINACY RELATIONS

The Heisenberg uncertainty principle is not normally considered a "quantum conundrum" yet it is still the topic of a great deal of theoretical analysis. As we noted elsewhere (Boughn & Reginatto 2018):

> Nevertheless, the question of how to relate Heisenberg's analysis to the formalism of quantum mechanics has never been completely settled and discussions about the meaning of and primacy of Heisenberg's uncertainty relations have continued unabated. These discussion often focus on one or more of the different expressions of the uncertainty principle such as: restrictions on the accuracy of simultaneous measurements of canonically conjugate quantities, e.g., $p$ and $q$; restrictions on the spread of individual measurements of conjugate quantities made on an ensemble of similarly prepared systems; restrictions on the physical compatibility of experimental arrangements for accurately measuring different observables; and the inevitable disturbance of a system due to its interaction with a measuring device.

Therefore, it is entirely appropriate to include the uncertainty principle in a discussion of the philosophical foundations of physics.

For me, Heisenberg's uncertainty principle was perhaps the most important piece of my education in introductory quantum mechanics. It is empirical in that it involves a (hypothetical) gamma ray microscope and, perhaps for that reason, it provided me with something tangible to help grasp the mysteries of quantum mechanics. This feeling was bolstered by the common (textbook) interpretation of the meaning of Heisenberg's principle, that one cannot simultaneously measure $p$ and $x$ without incurring errors in the measurements that satisfy an inequality. Unfortunately, any insight I seemed to acquire from Heisenberg's principle quickly evaporated. The standard expression of the uncertainty principle, $\Delta p \Delta x \geq \hbar/2$, was derived by Kennard four months after Heisenberg's paper. Kennard's derivation is straight from the formalism of wave mechanics and makes no reference to actual measurements of $p$ and $x$. The $\Delta$'s in the expression refer to the variances of **p** and **x** operating on a wave function $\Psi$, e.g., $(\Delta p)^2 = \int \Psi^* \boldsymbol{p}^2 \Psi dx - [\int \Psi^* \boldsymbol{p} \Psi dx]^2$, and so the uncertainty principle is simply a statement about the statistical dispersions about the expectation values of these operators relative to a given wave function whether or not any measurements are made. Furthermore, if measurements are made, there is no requirement that they be simultaneous only that they are made relative to the same quantum state $\Psi$. That is, the Heisenberg uncertainty principle, as stated above, is simply a statement about statistical dispersions of quantum operators with respect to quantum states and has nothing to do with observational or experimental uncertainty. You can imagine how this understanding served to undermine my hope for developing some experimental intuition about quantum mechanics.

---

[21] While the analysis in that paper was somewhat detailed, straightforward dimensional analysis gives the same result.

About 10 years ago I came upon an English translation of Heisenberg's 1927 paper (Wheeler and Zurek 1983) and read it with interest. If one considers Heisenberg's microscope not as a measuring apparatus but rather as a quantum state preparation device, it is straightforward to make a connection between this device and the statistical dispersions of quantum operators represented in the standard expression of the uncertainty principle (Ballentine 1970). However, Heisenberg's purpose was much more general than this and more in line with my initial reading of the uncertainty principle. His pragmatic point of view certainly resonated with me. Very early in his paper, Heisenberg noted:

> When one wants to be clear about what is to be understood by the words "position of the object", for example of the electron (relative to a given frame of reference), then one must specify definite experiments with whose help one plans to measure the "position of the electron"; otherwise, this word has no meaning.

Later in the paper, as noted in footnote 12, he cited an analogous case in relativity:

> It is natural in this respect to compare quantum theory with special relativity. According to relativity, the word "simultaneous" cannot be defined except through experiments in which the velocity of light enters in an essential way….We find a similar situation with

> the definition of the concepts of "position of an electron" and
> "velocity" in quantum theory. All experiments which we can use
> for the definition of these terms necessarily contain the uncertainty
> implied by equation (1) [$\delta p \delta x \sim h$].

These statements are an indication of why Heisenberg (at least the young Heisenberg) was labeled a logical positivist by some. They are certainly direct examples of the application of the *1st corollary* to my pragmatic philosophy presented in Section 2.

Heisenberg introduced a $\gamma$-ray microscope with which one can determine the position of an electron with arbitrary accuracy so long as the wavelength of the $\gamma$-ray is small enough. His argument was that it is only possible to determine the position to an accuracy on the order of the wavelength $\lambda$ of the $\gamma$-ray, i.e., $\delta x \sim \lambda$. In the process of making this measurement, the $\gamma$-ray imparts a momentum impulse to the electron proportional to the momentum of the $\gamma$-ray photon, which by the Einstein relation is $h/\lambda$. The unknown momentum disturbance to the electron will also be of this order so that the uncertainty in the electron's momentum is $\delta p \sim h/\lambda$ (assuming the initial momentum of the electron is known). From these two relations for $\delta x$ and $\delta p$ one concludes that $\delta p \delta x \sim h$, Heisenberg's indeterminacy relation.[22] This simple relation was the first and most important in his paper. To be sure, Heisenberg considered the indeterminacy relation to be a direct physical interpretation of the quantum mechanical commutation relation, $\boldsymbol{px} - \boldsymbol{xp} = i\hbar$, but he clearly considered it to be an empirical result and not just a relation following from the formalism of quantum mechanics.

In the abstract to the paper, Heisenberg declared, "This indeterminacy is the real basis for the occurrence of statistical relations in quantum mechanics." This also resonated with my initial hope that the uncertainty principle would help me grasp quantum mechanics. Near the end of his paper he states,

> We have not assumed that quantum theory— in opposition to
> classical theory— is an essentially statistical theory in the sense
> that only statistical conclusions can be drawn from precise initial
> data. ...what is wrong within the sharp formulation of the law of

---

[22] Bohr and others criticized Heisenberg for omitting the details of the optics of the microscope. When these are included the result, $\delta p \delta x \sim h$, is unchanged. I suspect that Heisenberg was satisfied with his original argument as he noted: "...one does not need to complain that the basic equation [$\delta p \delta x \sim h$] contains only qualitative predictions." A qualitative relation was all he sought.

> causality, "When we know the present precisely, we can predict the future," is not the conclusion but the assumption. Even in principle we cannot know the present in all detail. For that reason everything observed is a selection from a plenitude of possibilities and a limitation on what is possible in the future. As the statistical character of quantum theory is so closely linked to the inexactness of all perceptions, one might be led to the presumption that behind the perceived statistical world there still hides a "real" world in which causality holds. But such speculations seem to us, to say it explicitly, fruitless and senseless.

The pragmatic tone of this declaration is quite clear. Heisenberg didn't stop there. He opined, "Of course we would also like to be able to derive, if possible, the quantitative laws of quantum mechanics directly from the physical foundations— that is, essentially, from relation (1) $[\delta p \delta x \sim h]$" but then was forced to conclude that "We believe, rather, for the time being that the quantitative laws can be derived out of the physical foundations only by use of the principle of maximum simplicity." There were two recent attempts (Hall & Reginatto 2002; Boughn & Reginatto 2018) to fulfill Heisenberg's wish. I'll leave it to the reader to decide whether these were successful. (I think they were. ☺) In a similar vein, I recently penned an essay (Boughn 2018c and 2019c) that addresses the question, "What is it about our world that led us to a quantum mechanical model of it?"

3.3 COLLAPSE OF THE WAVE FUNCTION

*Collapse of the wave function* (*state reduction*) is often considered to be part of the standard interpretation of quantum mechanics. However, Bohr did not consider it so. Just how did this concept enter the quantum lexicon? Some credit (blame?) Heisenberg from comments he made in his seminal paper on the uncertainty principle (Heisenberg 1927). There, he notes that as a consequence of observing an atomic system "we will find the atom has jumped from the $n^{th}$ [superposition] state to the $m^{th}$ state with a probability…" and, in a discussion of an electron wave packet, he comments, "Thus every position determination *reduces* [my italics] the wavepacket…" Perhaps, the first careful discussion of state reduction appeared in von Neumann's 1932 influential text *The Mathematical Foundations of Quantum Mechanics* (von Neumann 1955) in which he gave a quantum mechanical description of measurements. He specified two processes in quantum mechanics: the discontinuous change of states that occurs as a result of a measurement (state reduction); and the unitary evolution of quantum states (e.g., the Schrödinger equation). Dirac was more explicit in his equally revered text, *The Principles of Quantum Mechanics,* (Dirac 1958)

> When we measure a real dynamical variable $\xi$, the disturbance involved in the act of measurement causes a jump in the state of the dynamical system. From physical continuity, if we make a second measurement of the same dynamical variable $\xi$ immediately after the first, the result of the second measurement must be the same as that of the first. Thus after the first measurement has been made, there is no indeterminacy in the result of the second. Hence, after the first measurement has been made, the system is in an eigenstate of the dynamical variable $\xi$, the eigenvalue it belongs to being equal to the result of the first measurement. This conclusion must still hold if the second measurement is not actually made. In this way we see that a measurement always causes the system to jump into an eigenstate of the dynamical variable that is being measured, the eigenvalue this eigenstate belongs to being equal to the result of the measurement.

This seems like a reasonable conclusion; so what is my problem with it? For one thing, measurements rarely proceed in such a fashion. After a measurement, the object that the wave function describes is usually nowhere to be found. Even a relatively simple experiment as provided by a Stern-Gerlach apparatus usually ends with a polarized atom striking a photographic emulsion (or the modern equivalent) after which there is no conceivable way to measure its state again. To be sure, in special cases the quantum system can be measured to be in a particular eigenstate and afterwards can be found in that same state. In those cases it seems as if the wave function might actually have collapsed.

So how would a pragmatist like Bohr explain the "apparent" collapse of the wave function? Before answering this question, it is necessary to address the question of what we mean by the term "wave function". By "mean" I'm referring to the empirical meaning that an experimentalist like me needs to know. It is here that Stapp's 1972 paper on the Copenhagen interpretation informs us. In Stapp's practical account of quantum theory, a system to be measured is first prepared according to a set of specifications, *A*, which are then transcribed into a wave function $\Psi_A(x)$, where *x* are the degrees of freedom of the system. The specifications *A* are "couched in a language that is meaningful to an engineer or laboratory technician", i.e., not in the language of quantum (or even classical) formalism.

Likewise, *B* are a set of specifications of the subsequent measurement and its possible results. These are transcribed into another wave function $\Psi_B(y)$, where *y* are the degrees of freedom of the measured system. How are the mappings of *A* and *B* to $\Psi_A(x)$ and $\Psi_B(y)$ effected? According to Stapp,

> …no one has yet made a qualitatively accurate theoretical description of a measuring device. Thus what experimentalists do, in practice, is to *calibrate* their devices…[then] with plausible assumptions…it is possible to build up a catalog of correspondences between what experimentalists do and see, and the wave functions of the prepared and measured systems. It is this body of accumulated empirical knowledge that bridges the gap between the operational specifications *A* and *B* and their mathematical images $\Psi_A$ and $\Psi_B$. Next a transition function $U(x; y)$ is constructed in accordance with certain theoretical rules…the 'transition amplitude' $\langle A|B \rangle \equiv \int \Psi_A(x) U(x; y) \Psi_B^* \, dxdy$ is computed. The predicted probability that a measurement performed in the manner specified by *B* will yield a result specified by *B*, if the preparation is performed in the manner specified by *A*, is given by $P(A, B) = |\langle A|B \rangle|^2$.

Of course, our use of quantum wave functions is not limited to the experimentalist's laboratory. One often assigns a wave function to systems and measurements that have been prepared *naturally*. Consider, for example, the case of the emission of 21 cm radiation from atomic hydrogen in the interstellar medium. In that case one specifies the initial upper ground state wave function, the subsequent lower ground state wave function, and the interaction that results in the emission of radiation. In this case, the interstellar environment takes the place of the experimentalist's laboratory. Nevertheless, the conditions of this environment provide the operational specifications that the astronomer must translate into quantum states. Also, it is unlikely that such translations could ever be made without the laboratory experiments that informed us about quantum phenomena.

What's my point here? It is that the quantum mechanical wave function is a theoretical construct that we invented to deal with our observations of physical phenomena. As such, it seems reasonable that we derive our understanding of it according to how we use the concept. This is precisely the point of the *2nd corollary* of my pragmatic philosophy. Stapp's (and Bohr's) pragmatic account of wave functions is intimately tied to state preparation and measurement, both of which are described in terms of operational specifications that lie wholly outside the formalism of quantum mechanics. Prior to the preparation of a system the wave function is not even defined and after it has been measured the wave function ceases to have a referent. To extrapolate the notion of the wave function to its pivotal roll in von Neumann's and Dirac's process of state reduction is an enormous leap that necessitates dropping all references to the operational specifications that give wave functions their meanings in the first place. One might be led to such an extrapolation by ascribing an ontic reality to the notion of wave function (despite its epistemic transitory nature). This is especially tempting when presented with some of the incredible successes of quantum theory. I was certainly enticed to do so when I first learned of the twelve decimal place agreement of the quantum electrodynamic prediction with the measured value of the g-factor of the electron.[23] However, one should be wary of deeming a single (or even several) high precision measurement(s) as strong evidence in this regard.[24]

We now return to the question of how a pragmatist like Bohr might explain the "apparent" collapse of the wave function. The Copenhagen interpretation's resolution of the problem (according to Stapp 1972) is straightforward. In the case that the quantum system is disrupted upon the completion of a measurement, the pre-measurement wave function simply ceases to have meaning and there is nothing that requires the notion of wave function collapse. In the special cases mentioned above, the measurement does not disrupt the quantum system but leaves it in a state with a well-defined wave function that can be used to predict a subsequent measurement. This is the only time that the notion of wave function collapse might make sense. However, in this case the pragmatist would simply point out that the original specifications of the quantum system must be replaced by new ones that include the results of the first measurement. "Then the original wave function will be naturally replaced by a new one, just as it would be in classical statistical theory" (Stapp 1972). Some physicists prefer to claim, euphemistically, that wave function collapse happens in the mind

---

[23] When I was a graduate student it was considered prescient that the acronym for quantum electrodynamics (QED) was the same as for the Latin phrase *quod erat demonstrandum*.

[24] Otherwise, one might argue that spectroscopic measurements of hydrogen provided incontrovertible proof of the Bohr-Sommerfeld model of the hydrogen atom. The precise (1 part in $10^{14}$) timing measurements of the binary pulsar might be argued to imply the correctness of general relativity (Hawking and Penrose 1996) even though the current consensus is that classical general relativity is not a fundamental theory. Another example is Maxwell's equations, which predict a Coulomb inverse square law that has been confirmed to a precision of one part in $10^{16}$ even though we know that quantum electrodynamics is the more fundamental theory (Williams, *et al.* 1971; Fulcher 1986).

of the observer but not for any physical reason. For example Hartle (1968) noted,

> The "reduction of the wave packet" does take place in the consciousness of the observer, not because of any unique physical process which takes place there, but only because the state is a construct of the observer and not an objective property of the physical system.

Even thought Bohr's pragmatic interpretation of quantum mechanics long ago dispensed of the notion of wave function collapse, to this day extensions of quantum mechanics that hypothesize collapse mechanisms are still be proffered by distinguished theorists (e.g., Penrose 1996; Adler and Bassi 2009). As you can imagine, until new experimental observations point in that direction, a pragmatist might not be interested in such endeavors.

## 3.4 SCHRÖDINGER'S CAT AND DECOHERENCE

An issue closely related to wave function collapse is the paradox of Schrödinger's cat (Schrödinger 1935). If quantum mechanics is, indeed, the fundamental theory of the physical world, then presumably macroscopic systems like a cat are also governed by the theory. In Schrödinger's 1935 gedanken experiment, a cat is penned up in a chamber with a radioactive substance that is monitored by a Geiger counter. The half-life of the substance is such that after one hour there is a 50% chance that a single radioactive atom will have decayed in which case the counter discharges. If so, then a relay releases a hammer that shatters a flask of hydrocyanic acid and the cat dies. If the counter does not discharge, the flask is not broken and the cat lives. After one hour, the wave function of the entire system expresses this situation by having equal parts of an alive cat/undecayed atom and a dead cat/decayed atom and thus the "indeterminacy originally restricted to the atomic domain becomes transformed into macroscopic indeterminacy"(Schrödinger 1935). When the box is opened and the cat plus atom system observed, the wave function collapses into one of these two states. This is absurd. Surely, the cat is either alive or dead before the box is opened even if we don't know which is the case. In addition, if the observer is also viewed as a quantum system by another observer, Wigner's friend (Wigner 1967), then the first observer is also in one of two quantum states and the result of her observation doesn't become definitive until a second observer observes her. This conundrum can be carried on *ad infinitum* with the introduction of third, fourth, fifth,…. observers.

Is it ever meaningful to describe the cat as in a superposition of alive and dead states? Decoherence theory is often invoked as a resolution. As we described it elsewhere (Boughn & Reginatto 2013)

> Decoherence theory is neither new physics nor a new interpretation of quantum mechanics; although, it is certainly relevant to questions of interpretation. In decoherence theory, the measuring apparatus and the environment with which it inevitably interacts are both treated as purely quantum mechanical systems. As a consequence of the interactions of the quantum system of interest with the measuring apparatus and it with its immediate environment, the three become entangled, i.e., strongly correlated with each other. All, or at least most, of the environmental quantum degrees of freedom are not observable (certainly, not observed) and, therefore, must be summed over to achieve a reduced state of the system plus apparatus. The net effect of the enormous number of environmental degrees of freedom is that off-diagonal terms of the reduced density matrix rapidly vanish, i.e., coherence between the different eigenstates of the system/apparatus is lost. Thus, decoherence theory demonstrates why it is that quantum coherence is seldom, if ever, observed at the classical (macroscopic) level.

Quantum decoherence allows for the quantum treatment of the cat yet demonstrates why the live cat/undecayed atom and dead cat/decayed atom portions of the combined wave function will not exhibit quantum interference. This conclusion is reached without invoking wave function collapse but the paradox is still not resolved. Before being observed the live cat and dead cat are still described by a quantum state, albeit a mixed state, which cannot be distinguished experimentally from the classical claim that there's a 50% chance the cat is alive and a 50% chance the cat is dead. This mixed quantum state would still require quantum state reduction if the state of the cat is ultimately determined to be alive or dead and this is tantamount to admitting that before being observed, the cat is still in a composite state, the situation that vexed Schrödinger.

While I find these sorts of analyses interesting, I long ago came to doubt that there was any need to resolve this quantum/classical dilemma. For me, Schrödinger's cat poses no paradox. I certainly assume that in principle the cat can be treated quantum mechanically, in which case the quantum state would consist of the superposition of a live and dead cat. As pointed out in Section 1, I'm sure that Bohr and Heisenberg would agree. Perhaps they would even agree, again in principle, that one could perform an interference experiment to reveal the presence of both the live and dead states. On the other hand, the difficulty of performing such an experiment would render bringing the dead back to life trivial by comparison. Finally, even if I were to find the decoherence argument compelling, I would characterize it as simply demonstrating the *consistency* of the quantum mechanical microscopic world with the classical macroscopic world rather than demonstrating how classical physics might *emerge* from quantum mechanics. The pragmatic resolution of the paradox is simply to point out that there is no paradox. The entangled wave function is not to be viewed as an ontic representation of the cat but rather as a tool that physicists use to prepare systems and predict the results of measurements. As argued in Section 3.3, wave function collapse is not physically meaningful.

3.5 THE EPR PARADOX, BELL'S THEOREM, AND QUANTUM NONLOCALITY[25]

In the 84 years since the 1935 Einstein-Podolsky-Rosen paper (EPR), physicists and philosophers have mused about what Einstein referred to as *spooky action at a distance* (Born 1971). Bell's 1964 analysis of EPR-type experiments has been cited more than 10,000 times and I suspect that most of the citations have occurred in the last decade or two. Many of these describe experiments involving entangled quantum states and, as an experimentalist, I have great admiration for much of this work. Many more are theoretical and philosophical papers trying to come to grips with what is often referred to as quantum nonlocality. Bell's conclusion was that any hidden variable theory designed to reproduce the predictions of quantum mechanics must necessarily be nonlocal and allow superluminal interactions. In his words (Bell 1964)

> In a theory in which parameters are added to quantum mechanics to determine the results of individual measurements, without changing the statistical predictions, there must be a mechanism whereby the setting of one measurement device can influence the reading of another instrument, however remote. Moreover, the signal involved must propagate instantaneously, so that such a theory could not be Lorentz invariant.

---

[25] I've written four papers bearing on these subjects (Boughn 2016, Boughn 2017, Boughn 2018a, Boughn & Reginatto 2019) and the reader is referred to these for more detailed analyses. The following is a truncated synthesis of those papers.

I have no problem with this conclusion. Unfortunately, Bell later expanded his analysis and mistakenly concluded that quantum mechanics itself is nonlocal and this led him to apprehend the "gross nonlocality of nature" Bell (1975). Many physicists and philosophers of science seem to harbor this same belief.

While the EPR paper is often credited with bringing the notion of entangled quantum states and action at a distance to the attention of the physics community, the major figures in physics at the time were already arguing about entanglement and Einstein had been bothered by action at a distance since the 1927 Solvay conference (Howard 2007). At Solvay, he used a single slit gedanken experiment to illustrate the action at a distance required by quantum mechanics. Consider the simpler case of diffraction by a small hole. The Schrödinger wave function of a particle emanating from the hole is essentially a spherical wave indicating that the particle can be detected at any point on a hemispherical screen with equal probability. Therefore, the detection of the particle at any point on the screen must cause the wave function to instantaneously vanish everywhere except at that point, hence, action at a distance. Einstein pointed out that this was not a problem if the wave function is taken to represent the behavior of an ensemble of particles; however, if it is to provide an objective description of a single particle, then it clearly violates special relativity (Howard 2007). He wasn't challenging the veracity of the predictions of quantum mechanics but rather its ontological interpretation.

As I mentioned in Section 2, I was mystified when I first read the EPR paper. The conclusion of EPR was not that quantum mechanics is nonlocal nor that objective reality does not exist but rather that quantum mechanics "does not provide a complete description of the physical reality". In fact, Poldolsky penned the paper and Einstein was not happy with it. Howard (2007) points out that in a letter to Schrödinger written a month after the EPR paper was published, Einstein chose to base his argument for incompleteness on what he termed the "separation principle" and continued to present this argument "in virtually all subsequent published and unpublished discussions of the problem". According to the separation principle, the *real state of affairs* in one part of space cannot be affected instantaneously or superluminally by events in a distant part of space. Suppose *AB* is the joint state of two systems, *A* and *B,* that interact and subsequently move away from each other to different locations. (Schrödinger (1935) would later introduce the term *entangled* to describe such a joint state.) In his letter to Schrödinger, Einstein explained (Howard 2007)

> After the collision, the real state of (*AB*) consists precisely of the real state *A* and the real state of *B*, which two states have nothing to do with one another. *The real state of B thus cannot depend upon the kind of measurement I carry out on A* [separation principle]. But then for the same state of *B* there are two (in general arbitrarily many) equally justified [wave functions] $\Psi_B$, which contradicts the hypothesis of a one-to-one or complete description of the real states.

His conclusion was that quantum mechanics cannot provide a complete description of reality. Note that Einstein's separation principle did not claim that a measurement of system *A* has no effect on the *result of any measurement* on system *B* but rather that a measurement of system *A* has no effect on the *real state* of system *B*. I'll elaborate on this distinction shortly. You can probably already guess the pragmatic resolution of the EPR paradox; however, it's instructive to examine the details of the argument.

Perhaps the quintessential example of an EPR-type entangled state is the one introduced by Bohm and Aharanov (1957) and subsequently used by Bell in his 1964 paper. It consists of the emission of two oppositely moving spin ½ particles in a singlet state, i.e., the spins of the two particles are precisely opposite each other but in an indeterminate direction. According to quantum mechanics, their combined wave function (quantum state) is given by

$$\Psi(1,2) = \frac{1}{\sqrt{2}}\{|1,\uparrow\rangle_z|2,\downarrow\rangle_z - |1,\downarrow\rangle_z|2,\uparrow\rangle_z\} \quad (1)$$

where ↑ and ↓ indicate the up and down *z* components of the spins of particles 1 and 2. Now suppose that the spin of particle 1 is measured with a Stern-Gerlach apparatus oriented in the $\hat{z}$ direction and is determined to be ↑. Then we know the *z* component of the spin of particle 2 will be ↓. That is, particle 2 can be characterized by the state $\Psi(2) = |2,\downarrow\rangle_z$. This perfect (anti)correlation is built into the two-particle system because they are in a singlet state. On the other hand, the original wave function can also be expressed in terms of the *x* components (or in any direction for that matter) of the spin,

$$\Psi(1,2) = \frac{1}{\sqrt{2}}\{|1,\uparrow\rangle_x|2,\downarrow\rangle_x - |1,\downarrow\rangle_x|2,\uparrow\rangle_x\}.$$

Then, if the spin of particle 1 is measured with a Stern-Gerlach apparatus oriented in the $\hat{x}$ direction and is determined to be ↑, the spin of particle 2 will be ↓, i.e., particle 2 is in the state $\Psi(2) = |2,\downarrow\rangle_x$, a state fundamentally distinct from $|2,\downarrow\rangle_z$. The immediate transition of particle 2 to either the state $|2,\downarrow\rangle_z$ or $|2,\downarrow\rangle_x$ is, of course, the process of state reduction discussed in Section 3.3 and is the basis for the claim of "action at a distance".[26] After Bell's papers, some physicists and philosophers embraced the nonlocality inherent in quantum mechanics and the concomitant notion of instantaneous action at a distance. Likewise, Einstein felt that quantum mechanics implied action at a distance and, therefore, violated his separation hypothesis. Consequently, he concluded that quantum mechanics failed to provide "a one-to-one or complete description of the real states"; ergo, quantum mechanics is incomplete. So Einstein and adherents of quantum nonlocality both ascribe to spooky action at a distance but while the latter embrace it, Einstein inferred from it that quantum mechanics does not provide a complete description of physical reality.

Now consider a more pragmatic perspective. For the entangled singlet state considered above, Einstein would conclude particle 2 has no unique (real) state. Let me push back on this conclusion. After their emission, the polarization of neither particle is known. Therefore, the two particles can be considered as two unpolarized particle beams. For particle 2 this can be represented by a 50% mixture of spin up, $|2,\uparrow\rangle$, and spin down, $|2,\downarrow\rangle$, states (in any direction), a so-called *mixed state*. A mixed state cannot be described by a pure quantum state (e.g., Schrödinger wave function or vector in Hilbert space) but is well described by its associated density matrix $\varrho$. The density matrix for the unpolarized spin ½ particle 2 is given by

$$\rho(2) = \tfrac{1}{2}|2,\uparrow\rangle\langle 2,\uparrow| + \tfrac{1}{2}|2,\downarrow\rangle\langle 2,\downarrow| \qquad (2)$$

with a similar expression for particle 1. This density matrix is sufficient to completely describe any possible (spin) measurement of particle 2. For example the expectation value (mean measurement) of any observable represented by an operator $A$ is given by $\langle A \rangle = tr(\rho A)$, i.e., the trace of the matrix $\rho A$ and, as always, the value of a single measurement is one of the eigenvalues of $A$. For the above expression this implies that a measurement of the spin component in any direction will yield ↑ for 50% of the measurements and ↓ for 50% of the measurements subject to the usual statistical fluctuations.

---

[26] As discussed above, Dirac (1958) and von Neumann (1955) both gave detailed treatments of state reduction; although, it's not clear to what extent they ascribed to the notion of action at a distance.

This completely characterizes the spin measurements of either particle with no mention of what measurements are made on the other particle. That is, the *result* of the measurement made on particle 2 is completely independent of any measurement carried out on particle 1.[27] I would call this the experimentalist's or pragmatist's separation hypothesis in contrast to Einstein's version that the *real state* of particle 2 must be independent of any measurement carried out on particle 1. I know of no observation or prediction of standard quantum mechanics that violates the pragmatist's separation hypothesis. As for Einstein's separation hypothesis, a pragmatist would probably object to the notion of a *real state* or at least defer comment until an acceptable definition of "real state" is offered.[28]

On might object that Eq. 2 was pulled out of thin air and suggest that a proper quantum mechanical treatment might provide a more complete description of the individual particle states. Not so. Consider the density matrix of the pure entangled state in Eq. 1. It is given by $\rho(1,2) = |\Psi(1,2)\rangle\langle\Psi(1,2)|$ and is in every way equivalent to the entangled state $\Psi(1,2)$ itself. So how do we conjure a single particle 2 state from the entangled state? The standard method is to marginalize the two-particle density matrix over the possible states of particle 1. This is accomplished by taking the *partial trace* of density matrix $\rho(1,2)$ over a basis of system 1, i.e.,

$$\rho(2) = \langle 1,\uparrow| \rho(1,2)|1,\uparrow\rangle + \langle 1,\downarrow| \rho(1,2)|1,\downarrow\rangle.$$

This is precisely the density matrix of Eq. 2. That is, Eq. 2 describes the complete quantum state of particle 2 and characterizes any measurement made on that particle, of course subject to the statistical nature of all quantum mechanical predictions.

Okay, so far so good. Nevertheless, you may object that these single particle mixed states are mute on the correlations of measurements made on the two states. Fair enough, but now you are asking a different question. That is, after making these independent measurements, what are the correlations between them? To answer this question, to be sure, one needs the pure entangled state of Eq. 1 but this expression in no way implies that the measurements made on particle 1 have any effect whatsoever on the measurements made on particle 2. One might exhort, as did John Bell (1981), "The scientific attitude is that correlations cry out for explanation." Well, maybe but as we often have to remind our students, correlation does not necessarily imply causation and in any case one can certainly trace the source of the correlation back to the interaction between the two particles prior to their emission.

---

[27] After all, the measurement made on particle 2 could be made in advance of even selecting the type of measurement to be made on particle 1.

[28] I posed the question "What are Physical States?" in a recent essay (Boughn 2019b).

There is a simple classical analog that emphasizes this point. Suppose one randomly places either a white ball or a black ball in a box and then places the remaining ball in another box. Now the two boxes are closed and sent off in opposite directions where they will encounter observers who open the boxes. It's clear each observer has a 50/50 chance of finding a white ball and a 50/50 chance of finding a black ball and that completely characterizes the state of the unopened box for each individual observer. This can be confirmed by compiling the data from multiple (random) trials of the experiment. Yet it is also absolutely clear that the findings of the two observers will be completely correlated; if the observer of box 1 finds a white ball, the observer of box 2 will necessarily find a black ball and vice versa. However, we would never claim that the act of observing a white ball in box 1 *causes* a black ball to appear in box 2. Of course, quantum entanglement is a much richer phenomenon than classical entanglement and exhibits all the aspects of quantum interference with which we are familiar. But this results from quantum superposition that lies at the heart of quantum mechanics, and has little to do with the notion of entanglement and certainly does not provide evidence of nonlocality and action at a distance.

Still, after we observe particle 1 and determine the $z$ component of its spin to be ↑, aren't we justified in identifying the state of particle 2 to be $\Psi(2) = |2,\downarrow\rangle_z$? In which case, didn't we cause its quantum state to collapse from its original mixed state in Eq. 2 to $|2,\downarrow\rangle_z$? In a sense, the answer to this question is 'yes' but only in the sense of the pragmatic meaning of term "quantum state" as discussed in Section 3.3. So after the measurement of particle 1, what is the quantum state of particle 2? Is it still the density matrix $\rho(2)$ of Eq. 2 or is it the wave function $|2,\downarrow\rangle_z$? The premise of this question is wrong. The quantum state is not an "it" in this sense. The only "it" is the quantum system, the spin ½ particle. The quantum state is not a physical object but rather a description of a physical object (Boughn 2019b), a transcription arising from Stapp's "set of specifications". For the physicist who has just observed particle 1 to be in the ↑ state, the set of specifications leads to $\Psi(2) = |2,\downarrow\rangle_z$ while for the future observer of particle 2, the available set of specification leads to the density matrix $\rho(2)$ of Eq. 2. If the result of the observation of particle 1 is communicated (subluminally, of course) to the observer of particle 2, then the original specifications of the quantum system must be replaced by new ones that include the result of that measurement. "Then the original wave function [or density matrix in this case] will be naturally replaced by a new one, just as it would be in classical statistical theory"(Stapp 1972). So for the pragmatist, the EPR paradox is no paradox at all but rather is simply another example of the wonders of standard quantum mechanics.

Einstein contended that quantum mechanics implies "spooky action at a distance"; physicists and philosophers who accept quantum nonlocality are in agreement. Their disagreement is a metaphysical one. Einstein didn't like this aspect of quantum mechanics whereas the latter group embraces it. So where do they go wrong? As I alluded to above, there is often a tendency to identify theoretical constructs of highly successful models with reality itself. One can argue that this is as much the case with classical mechanics as with quantum mechanics. The "particle" construct of the former is considered to be real in the same sense as the "wave function" construct of the latter. Once one is lead down this path, it is inevitable to conclude that spooky action at a distance occurs in nature. On the other hand, if one eschews the ontological interpretation of the wave function, then "action at a distance" is, at best, descriptive of how the mathematical formalism of quantum mechanics is applied. One must be extremely wary of extending such inferences to physical reality itself and, in fact, doing so is a violation of the *2$^{nd}$ corollary* to my pragmatic philosophy. I've always thought the proofs that standard quantum mechanics forbids sending superluminal signals should have disabused nonlocality advocates of their claims of action at a distance. Once one attaches ontic significance to theoretical constructs, like the wave function, such metaphysical conclusions are inevitable.

3.6 DOES GOD PLAY DICE?

A common thread running through the previous four quantum conundrums is the statistical nature of quantum mechanics. Einstein was quite vocal on this count as he famously declared, "I…am convinced that *He* [God] is not playing at dice." (Born 1971)

I confess that the statistical aspect of quantum mechanics also greatly bothered me in my younger days. I usually expressed my dismay with something like

> The deterministic Schrödinger equation governs the evolution of the wave function (probability amplitude) but as far as I can discern, quantum theory proper is silent on the outcome of any particular measurement. According to quantum mechanics, nothing ever happens!

As I matured as an experimental physicist, I became acutely aware of the fact that everything I did was intimately connected to statistics and conclusions nearly always involved probabilistic statements. This was so whether I was dealing with quantum mechanical or classical measurements. So, as far as I'm concerned, classical physics is also a statistical affair (Boughn & Reginatto 2019).

Heisenberg felt that his indeterminacy relation was "the real basis for the occurrence of statistical relations in quantum mechanics" (see Section 3.2) and to a certain extent I agree with his conclusion. But there is more to it than this. First, what are the statistical relations in classical mechanics and from where do they arise? Compare the uncertainty relation in quantum mechanics, $\Delta p \Delta x \geq \hbar/2$, with a corresponding relation in classical mechanics, $\Delta p \Delta x \geq 0$. The minimum uncertainties in the two cases, $\hbar/2$ and $0$, reflect differences in the quantum and the classical formalisms. Such was Heisenberg's reasoning. What about the uncertainties in excess of $\hbar/2$ and $0$? For the quantum case the answer to this question is nuanced and I'll return to it shortly. In classical mechanics uncertainties are usually ascribed to "noise" and delegated to experimentalists for clarification. However, this doesn't have to be the case. There are formulations of classical mechanics that allow one to include uncertainties in the formalism, just as in quantum mechanics.

A particularly powerful description of classical mechanics can be formulated using ensembles on configuration space (Hall & Reginatto 2005, 2016). A Hamiltonian formulation for a classical particle of mass $m$ under the influence of a potential $V$ leads to the equations of motion,

$$\frac{\partial P}{\partial t} + \frac{\nabla \cdot (P \nabla S)}{m} = 0 \quad \text{and} \tag{1}$$

$$\frac{\partial S}{\partial t} + \frac{|\nabla S|^2}{2m} + V = 0 \tag{2}$$

where $P(x,t)$ is the probability density for the location of the particle and $S(x,t)$ is field canonically conjugate to $P$. The first equation is the probability continuity equation and the second is the classical Hamilton-Jacobi equation. The quantity $P\nabla S$ is interpreted as the local momentum density. If the initial values of $P$ and $S$ are chosen so as to represent the uncertainty of the initial state of the particle, then the subsequent uncertainties are determined from the equations of motion. In this formalism classical uncertainty, i.e., noise, can be included in the formalism just as it is in quantum mechanics. The nonlinear Hamilton-Jacobi equations are certainly unwieldy and less useful than Newton's laws. However they are, in some sense, more general than Newton's equations, which in any case are recovered by specifying $P$ to be a Dirac delta function.

It is interesting that the H-J formalism is an alternative with which one can express standard quantum mechanics, as was pointed out by Madelung in the early days of quantum mechanics (Madelung 1927). This is accomplished by adding a quantum term to the Hamilton-Jacobi equation,

$$\frac{\partial S}{\partial t} + \frac{|\nabla S|^2}{2m} + V - \frac{\hbar^2}{2m}\frac{\nabla^2 P^{1/2}}{P^{1/2}} = 0 \qquad (3)$$

and retaining the continuity equation (1). Equations (1) and (3) are equivalent to the Schrödinger equation where the wave function $\Psi$ is related to $P$ and $S$ by $\Psi = \sqrt{P}e^{iS/\hbar}$ (Hall & Reginatto 2005, 2016). The Hamilton-Jacobi formulation of classical and quantum mechanics also points to the importance of the quantum of action in the difference between the two. If one sets $\hbar = 0$, the quantum formulation reduces to the probabilistic formulation of classical mechanics.

To be sure, quantum mechanics is quite different from classical mechanics. The addition of the quantum term in Equation (3) has enormous consequence. In addition to the transition of $\Delta p\Delta x \geq 0$ to $\Delta p\Delta x \geq \hbar/2$, Equation (3) gives us wave/particle duality and all of the wonders associated with quantum interference. The point is that these follow from Planck's quantum of action and not from an inherent statistical nature of the theory.

In the above discussion I've implied there is an analogy between the quantum uncertainty in excess of $\Delta p\Delta x = \hbar/2$ and classical uncertainties, i.e., noise. In what sense is this so? Excess quantum uncertainty occurs for most systems in pure quantum states.[29] How is this analogous to classical noise? Recall Stapp's (1972) prescription for transcribing a set of physical specifications into a wave function (pure quantum state),

> … what experimentalists do, in practice, is to *calibrate* their devices… [then] with plausible assumptions…it is possible to build up a catalog of correspondences between what experimentalists do and see, and the wave functions of the prepared and measured systems.

The resulting wave function then leads directly to the excess uncertainties in Heisenberg's principle. On the other hand, the same set of procedures leads classical physicists to arrive at noise estimates: careful calibration of devices; catalogs of the results of many measurements; subsequent estimates of uncertainties in these measurements; and the transcription of these uncertainties into the appropriate $P$ and $S$. In the case of mixed quantum states, the analogy with classical uncertainty is more direct. Consider the unpolarized beam of spin ½ particles introduced in Section 3.5. The quantum state is represented by the density matrix $\rho = p|\uparrow\rangle\langle\uparrow| + p|\downarrow\rangle\langle\downarrow|$ where $p = 0.5$ is the probability for each of the two spin states and is, in every sense, equivalent to the classical probability of the spin components.

So the pragmatist's answer to Einstein is that all of physics has probabilistic aspects. Whether this is dealt with in the theoretical formalism or characterized as noise is irrelevant. After all, the models we choose to represent the world and make sense of our experiences are, in the end, human inventions.

3.7 GIBBS PARADOX AND THE SECOND LAW OF THERMODYNAMICS

So far, 5 out of the 6 conundrums I have discussed are quantum mechanical. Let me end with a thoroughly classical paradox from 1875 that was posed by J. Willard Gibbs, one of the founders (along with Maxwell and Boltzmann) of *statistical mechanics*. The Gibbs paradox involves the entropy of mixing and seemingly results in a violation of the *second law of thermodynamics*. Suppose there are two volumes, $V_1$ and $V_2$, of ideal gases separated by a diaphragm, with $N_1$ and $N_2$ molecules respectively. The two volumes of gases are presumed to be chemically distinct. Also assume that the two volumes of gas have the same temperature $T$ and pressure $P$. From classical thermodynamics (and kinetic theory) we can express the entropies of the two volumes as

$$S_{1,2} = N_{1,2} k \left[ log(V_{1,2}) + \frac{3}{2} log(T) \right] + const.,$$

where $k$ is Boltzmann's constant. Now remove the diaphragm and allow the two gases to mix. The entropy of this combined mixture is

$$S_{tot} = (N_1 + N_2) k \left[ log(V_1 + V_2) + \frac{3}{2} log(T) \right] + const.$$

---

[29] For the single particle case, only a minimum Gaussian wave packet results in no excess noise, i.e., $\Delta p \Delta x = \hbar/2$, and even in that case the system immediately evolves so as $\Delta p \Delta x > \hbar/2$.

Therefore the change in the entropy, the entropy of mixing, is

$$\Delta S = S_{tot} - S_1 - S_2 = (N_1 + N_2) k \, log(V_1 + V_2) - N_1 k \, log(V_1) - N_2 k \, log(V_2).$$

In the case that $N_1 = N_2 = N/2$ and consequently $V_1 = V_2 = V/2$, this becomes $\Delta S = Nk \, log(2)$, the entropy of mixing. The quantity $T\Delta S$ represents both the useful work that can be extracted from the system if the mixing is accomplished in a reversible manner as well as the minimum work required to separate the two mixed gases back to their original states. I leave it as an exercise for the reader to demonstrate that both of these processes can be realized if one is in possession of two semipermeable membranes, one of which allows only gas molecules of volume $V_1$ to pass through and the other only allows molecules of $V_2$ to pass.[30] It should be pointed out that "returning the gases to their original state" means only that all the molecules of type 1 are returned to $V_1$ and all the molecules of type 2 are returned to $V_2$, not that their original microstates, positions and momenta, are restored.

Now to the "paradox". Suppose the two volumes are filled with the same gas. When the diaphragm is removed, the gases are again mixed and because there is no identification of types of gas in the above relations, the entropy of mixing is still given by $\Delta S = Nk \, log(2)$. Now, simply replace the diaphragm and we immediately have the original state; all the macroscopic thermodynamic variables are the same as is the chemical composition of the two volumes. Therefore, entropy of the systems seems to have decreased by $Nk \, log(2)$, with no work nor any change in the entropy of the external world. This is a clear violation of the 2nd law of thermodynamics. It's interesting that Gibbs never referred to this situation as a paradox. We'll return to this shortly.

The standard resolution of the paradox follows from the statistical mechanical expression for the entropy in terms of the number of microscopic states, $\Omega$, available to the macroscopic thermodynamic state, i.e., $S = k \, log(\Omega)$. The argument is that interchanging any two molecules in the microscopic state of a system renders no change whatsoever in the macroscopic thermodynamics state.[31] Because there are $N!$ ways of rearranging $N$ particles of the same type, one must reduce $\Omega$ by this factor to take into account the over-counting of states. Therefore, the expression for entropy must be reduced by $-k \, log(N!)$. By Stirling's approximation, this is $\approx -k \, [N \, log(N) - N]$ so

$$S_{1,2} = N_{1,2} k \left[ log(V_{1,2}) + \frac{3}{2} log(T) \right] - k \left[ N_{1,2} \, log(N_{1,2}) - N_{1,2} \right] + const.$$

---

[30] In the latter process, it is possible to unmix the gases with the minimum possible work, $W = T\Delta S$.

Likewise, the total entropy after mixing must be reduced by $-k \, log([N_1 + N_2]!)$. Again using Stirling's approximation we have

$$S_{tot} = (N_1 + N_2) k \left[ log(V_1 + V_2) + \frac{3}{2} log(T) \right] - k [(N_1 + N_2) log(N_1 + N_2) - (N_1 + N_2)] + const.$$

Using the ideal gas law, $PV = NkT$, we conclude that $\Delta S = S_{tot} - S_1 - S_2 = 0$, as desired. The entropy of mixing has vanished. In the case that $N_1$ and $N_2$ represent different chemical species, there are only $N_1! \, N_2!$ ways of rearranging particles of the same type so that the entropy of the mixed gas must be reduced by $-k \, log(N_1! \, N_2!)$. I leave it as an exercise for the reader to show that in this case the entropy of mixing is the same as previously computed, that is, for equal volumes of gas $\Delta S = Nk \, log(2)$.

There are two bothersome aspects of the above argument. Just how alike must the two types of gas molecules be in order to justify replacing $N_1! \, N_2!$ with $(N_1 + N_2)!$ in the Boltzmann relation $S = k \, log \left( \Omega / N_1! \, N_2! \right)$ and why must we divide by these configurations in the first place, i.e., why aren't these configurations to be counted in the available microscopic states? The standard textbook response to these questions invokes the quantum nature of matter. For particles of the same type, there is no absolutely no distinction between a particular multi-particle quantum state and the quantum state with any two of the particles interchanged. A statement like the following usually accompanies this argument:

> The Gibbs paradox thus foreshadowed already in the
> last century conceptual difficulties that were resolved
> satisfactorily only by the advent of quantum mechanics. (Reif 1965)

---

[31] This is analogous to the point made in arguing that replacing the diaphragm for a single component ideal gas restores the system to its original thermodynamic state.

While this explanation may seem like a *fait accompli*, there's something a bit strange about it. How is it that a conundrum in the formalism of one physical model, classical thermodynamics and statistical mechanics, is resolved by the formalism of another physical model, quantum mechanics? One may get the mistaken impression that the Gibb's paradox is somehow intimately related to quantum phenomena. It's hard for me to believe that the concept of entropy in classical physics could possibly presage quantum mechanics.

So it seems that entropy has two different descriptions. Which is right? Does it follow the Clausius prescription, $S = \int \frac{dQ}{T}$, or the Boltzmann prescription, $S = k \, log(\Omega)$? The premise of this question is wrong in the same sense as in section 3.5. Entropy is not an "it" in this sense. The only "it" is the thermodynamic system. As noted in Section 3.5 with regard to the quantum state, entropy is not a physical property of a thermodynamic system but rather a description of that system. The only sensible resolution must come from considering how physicists use this concept, the *2$^{nd}$ corollary* to my pragmatic philosophy of physics (see Section 2).

The most direct and classical extrication from Gibbs' paradox comes from the realization that entropy is not an ontic property of matter but rather is a concept that derives its meaning from the manner in which physicists use it. In classical thermodynamics, the meaning of the entropy of mixing is derived, in part, from the thesis that the quantity $T\Delta S$ represents the useful work that can be extracted from the system if the mixing is accomplished in a reversible (isothermal) manner and as the minimum amount of work required to separate the two mixed gases back to their original states. If two different gases are so similar that there is no known process by which useful work can be extracted from the process of mixing as well as no known thermodynamics process by which the gases can be subsequently separated, then for all practical circumstances the two gases can be considered indistinguishable. Then in the Clausius prescription one can simply declare there is no entropy of mixing, $\Delta S_{mixing} = 0$, and in the Boltzmann prescription that the number of microstates is given by $\Omega \div (N_1 + N_2)!$. There is no necessity to appeal to quantum mechanics and the identicality of quantum particles. Gibb's fully understood this point as early as 1874; however, his explanation seems to have been largely forgotten by subsequent textbook authors ( Jaynes 1992). As Jaynes puts it

> The difference in $\Delta S$ on mixing of like and unlike gases can seem

> paradoxical only to one who supposes, erroneously, that entropy is a property of the microstate… thermodynamics has a greater flexibility in useful applications than is generally recognized. The experimenter is at liberty to choose his macrovariables as he wishes; whenever he chooses a set within which there are experimentally reproducible connections like an equation of state, the entropy appropriate to the chosen set will satisfy a second law that correctly accounts for the macroscopic observations that the experimenter can make by manipulating the macrovariables within his set.

The above discussion is by no means intended to put an end to the investigation of entropy, which is one of the most fascinating and useful concepts in both classical and quantum physics. Many eminent physicists, including Gibbs, Boltzmann, Planck, Einstein, Schrödinger, Pauli, and others have made important contributions elucidating the concept of entropy, and this tradition continues today (e.g. Lieb and Yngvason 2000). These efforts are part of the process of creating useful stories that help us make sense of the physical world. There needn't be a single story about entropy. In fact, this is a common theme in thermodynamics. It is the flexibility of thermodynamics that is the source of its strength in dealing with a wide variety of physical phenomena.

## 4. FINAL REMARKS

Now that I have outlined my pragmatic philosophy of the foundations of physics, it is incumbent upon me to evaluate its utility, that is, if I am to be true to its pragmatic label. The examples recounted in Sections 1 and 3 were intended to do just that. In short, they were proffered to demonstrate how various paradoxes and conundrums in physics have been perpetuated primarily by attributing ontic existence to theoretical constructs as if they were components of a physical reality. My pragmatic philosophy is intended to point this out and then dismiss them as faux problems that, in principle, will never be resolved. Of course, the two classical paradoxes, the twin paradox and the Gibbs paradox, have been dismissed long ago; however, even about these, misleading pronouncements continue to be made. On the other hand, the quantum conundrums, especially the quantum/classical divide, wave function collapse, and the spooky action at a distance of entangled states, have remained unresolved after nearly a century. This ought to be an indication that there is something seriously wrong with the theoretical interpretations from which they emerge as I have endeavored to demonstrate.

On the other hand, it can be argued that attempts to resolve apparent paradoxes have led to important advances in physics or at least to important advances in understanding physical models. As I have already pointed out, the Gibbs paradox has stimulated important work on entropy that has continued to this day. Even the continuing discussions of the twin paradox have contributed to the pedagogy of special relativity. Regarding the quantum arena, many have attributed John Bell's 1964 paper with initiating both experimental and theoretical advances in quantum entanglement and thereby to progress in the new fields of quantum information and quantum computing, although I'm less sanguine about this claim. In 1957, Hugh Everett introduced the "many worlds" interpretation of quantum mechanics[32] precisely in order to solve two of the primary aspects of the measurement problem, the classical/quantum divide and the collapse of the wave function. While he was motivated by, what I consider to be, non-existent problems[33], many people credit Everett with inspiring them to pursue research in new areas of physics, including decoherence theory, quantum information, and the application of quantum mechanics to cosmology (Bryne 2010). While I doubt many would declare that the measurement problem in quantum mechanics has been successfully resolved[34], I can't argue with these physicists or even pretend to know what motivates their endeavors, theoretical or experimental. A discussion of the source of scientific creativity is certainly beyond my poor powers and I won't attempt to address this topic here.

There are several topics I have avoided even though they have had an exaggerated impact on fundamental physics in the last few decades. I use the term "exaggerated" because of the dearth of experimental/observational evidence associated with them and until there is more evidence, there is little a pragmatist can say about them. These include dark matter and dark energy, inflation, and string theory.

The notion of dark matter has been with us for nearly a century. It is (presumably) a new kind of matter that is known only through its gravitational attraction to ordinary matter and to itself but is otherwise invisible. The mounting evidence for its existence has increased dramatically over the last 40 years, but as of yet there has been no direct detection. Even so, a great deal of progress has been made in ruling out various classes of models of dark matter. Is the claim of the existence of dark matter true, i.e., useful? I would say the answer is 'yes' in the sense that once introduced, it has proved to be a useful concept in the description of the gravitational clustering on many different scales. For example, there is evidence for unseen sources of gravity on the scales of individual galaxies, clusters of galaxies, and even on the large-scale structure of the universe.

---

[32] To be sure, it was Bryce DeWitt and Neil Graham (1973) who coined the term "many worlds". Everett didn't subscribe to this description of his "universal wave function".

[33] While I am not particularly enamored with Everett's theory of the universal wave function, I've concluded (Boughn 2018b) that, contrary to their own assessments, Everett's and Bohr's interpretations have a great deal in common.

[34] Recall, however, that I don't consider the problems of measurement in quantum mechanics even need to be resolved (Boughn & Reginatto 2013).

Dark energy is a generalization of the cosmological constant introduced by Einstein a century ago. Like dark matter, it is an "energy" only known through its gravitational effect but, in this case, its only consequence is the acceleration of the expansion of the universe. There have, indeed, been proffered many theoretical models of dark matter and dark energy. Some of these models seem to fit, at least qualitatively, into other fundamental physical theories. However, until a specific model of dark matter or dark energy has been verified by observations other than its gravitational attraction or the strength of that attraction has been shown to follow directly from some other physical theory, there is little a pragmatist like me has to offer.[35]

Inflation is a model that predicts the exponential expansion of the early universe. It explains both why the current 3-dimensional geometry of the universe is approximately Euclidean and, in addition, why the universe is so uniform on large scales. Like models of dark matter and dark energy, there are qualitative reasons why we might expect inflation to follow from other physical theories. So far there are no observations, other than the large-scale structure of the universe, that confirm or refute inflation. Worse yet, there are so many models of inflation that it seems certain any set of observations are likely to agree with one or more of those models. After one of the models has been identified, then it might be possible to help confirm it with new observations. Until then, like dark matter and dark energy, there is little a pragmatist like me has to offer.

My knowledge of string theory is nearly nonexistent so my comments regarding it are offered only from afar. During a talk at NYU, I once heard Gordon Kane pronounce that the mathematics of string theory is so powerful and so general that it can describe nearly any phenomenon and then proceeded to say that quantum gravity seemed so natural within its context that string theory was surely the true description of quantum gravity.[36] Hmmm…. There also seems to be a dearth of specific calculations relevant to empirical results so that it's difficult for a pragmatist to evaluate the truth, i.e., the utility, of string theory. In fact, I've often heard theorists proclaim that we still don't know how to construct a complete theory. Still others claim that there are $10^{500}$ or more versions of string theory. On the other hand, there are a great many theorists who extol the virtues of the theory. How so? Certainly the mathematical beauty of the theory motivates some as well as the possibility of unifying all the fundamental theories of physics into a single "theory of everything". Richard Dawid's recent book (2013), *String Theory and the Scientific Method*, argues for a new definition of scientific progress on the basis of purely theoretical notions. His premise relies heavily on an ontological view of physical theories so, again, a pragmatist like me has little to offer on this account.

---

[35] I have personally embarked on several unsuccessful attempts to directly detect dark matter and (with Rob Crittenden) succeeded in finding additional evidence for the gravitational effect of dark energy in the context of the integrated Sachs-Wolfe effect (Boughn & Crittenden 2004).

Mathematics certainly figures heavily in any discussion of string theory as well as to the general view of physical theories as espoused by Dirac and Tegmark (see Section 2). You probably will have noticed that I haven't mentioned where mathematics fits into my pragmatic philosophy. My knowledge of the fundamentals of mathematics is as meager as of string theory so I can say very little about it. However, I would hope that one might be able to take a pragmatic view of mathematics, in which case truth would be judged by utility. I can imagine that the usefulness of a certain area of mathematics can be judged by the degree with which it can be applied to other areas of mathematics or perhaps to the extent that it can help unify the content within a given area. Still, as a pragmatist, I would hope that its utility can also be judged by how useful it is in other areas of human activity, physics and beyond. Certainly mathematics, up till now, has proven to be extraordinarily useful to human endeavors. Otherwise, mathematics might be relegated to the status of a game (albeit an incredibly complex game) like *go* or chess.

Finally, I can't resist pointing out the irony presented by my pragmatic philosophy of 44physics to a self-avowed atheist like myself. If, indeed, physical theories are merely stories, all be they extremely intricate stories, about our experiences, then they would seem to be on the same footing as the stories of religion. Although I have no problem with scientists (e.g., Richard Dawkins and Lawrence Krauss) who criticize religion, especially organized religion, I've long ago come to the conclusion that trying to prove whether or not *god* exists is a useless endeavor.[37] My only recourse is to ask to what extent religious stories are true, i.e., useful. For me, the answer is not at all.[38]

---

[36] Among my philosophically motivated endeavors, I've written a couple of papers suggesting that perhaps there is no need for a quantum theory of gravity (Rothman & Boughn, 2006; Boughn 2009).

## Acknowledgements

A great many people have contributed to the views expressed in this paper and I would like to acknowledge them; although, of course, they should not be blamed for any foolish proclamations I might have made. Freeman Dyson and Marcel Reginatto are at the top of the list because of their tolerance of my ruminations over the last 10 to 15 years. Others who have similarly indulged me over the years are, in chronological order, Bob Taber, Jim Hollenhorst, Hojung Paik, Peter Pesic, Jeff Kuhn, Juan Uson, Jim Peebles, Mike Seldner, Ed Groth, Dave Wilkinson, Lyman Page, Norm Jarosik, Tony Rothman, Eliot Lieb, and I'm sure many others whom I've neglected to mention.## References

Adler, S., and Bassi, A., 2009, "Is Quantum Theory Exact?", *Science* **325**, 275.

Ballentine, L., 1970, "The statistical interpretation of Quantum mechanics", *Rev. Mod. Phys.* **42**, 358–381.

Bell, J., 1964, "On the Einstein Podolsky Rosen Paradox", *Physics* **1**, 195–200.---

[37] At the age of 13, I managed to construct a proof for the existence of god (what philosophers refer to as the cosmological proof). During my high school years I became increasingly dubious about my proof and then, in my first semester of college, had it totally destroyed by Ockham's razor in a philosophy course taught by Carl Hempel.

[38] I once got in an argument with a postmodern historian who insisted that the big bang theory was merely the latest in a long line of creation myths. I argued that the big bang was more an observation than a theory (myth). After going back an forth with no resolution, I ended the conversation with "Okay I agree that scientific theories are myths but the next time I fly to California to visit my daughter, I think I'll fly in a plane designed by an aeronautical engineer, not one conceived by a postmodernist." Utility indeed!